\documentclass[fleqn,usenatbib]{mnras}

\usepackage{newtxtext,newtxmath}

\usepackage[T1]{fontenc}
\usepackage[most]{tcolorbox}
\usepackage[shortlabels]{enumitem}

\DeclareRobustCommand{\VAN}[3]{#2}
\let\VANthebibliography\thebibliography
\def\thebibliography{\DeclareRobustCommand{\VAN}[3]{##3}\VANthebibliography}

\usepackage{graphicx}	
\usepackage{amsmath}	

\title[]{Probing viscosity of the intracluster medium using ram-pressure stripping}

\author[]{
Yung-Hsuan Tseng,$^{1}$\thanks{E-mail:jacinda@gapp.nthu.edu.tw}
Hsiang-Yi Karen Yang,$^{2,3,4}$\thanks{E-mail: hyang@phys.nthu.edu.tw}
Ryan Farber,$^{5}$
Mateusz Ruszkowski$^{6}$
\\
$^{1}$Department of Physics, National Tsing Hua University, Hsinchu 30013, Taiwan (R.O.C.)\\
$^{2}$Institute of Astronomy, National Tsing Hua University, Hsinchu 30013, Taiwan (R.O.C.)\\
$^{3}$Center for Informatics and Computation in Astronomy, National Tsing Hua University, Hsinchu 30013, Taiwan (R.O.C.)\\
$^{4}$Physics Division, National Center for Theoretical Sciences, Taipei 106017, Taiwan (R.O.C.)\\
$^{5}$Department of Physics, Purdue University Fort Wayne, Fort Wayne, IN 46805, USA\\
$^{6}$Department of Astronomy, University of Michigan, Ann Arbor, MI 48109, USA
}

\date{Accepted XXX. Received YYY; in original form ZZZ}

\pubyear{2024}

\begin{document}
\label{firstpage}
\pagerange{\pageref{firstpage}--\pageref{lastpage}}
\maketitle

\begin{abstract}
Galaxies falling into galaxy clusters can leave imprints on both the corona of galaxies and the intracluster medium (ICM) of galaxy clusters. Throughout this infall process, the galaxy’s atmosphere is subjected to ram pressure from a headwind, leading to the stripping morphology observed in its tail. The morphological evolution is affected by the properties of the surrounding ICM such as magnetic fields and viscosity. In this Letter, we perform 3D Braginskii-magnetohydrodynamic simulations using the FLASH code with varied ICM viscosity models. 
Specifically, we explore four models: an inviscid case, unsuppressed isotropic viscosity, unsuppressed anisotropic viscosity, and anisotropic viscosity suppressed by plasma instabilities. Our findings indicate that the isotropic viscosity case effectively suppresses hydrodynamic instabilities and shows strong viscous heating and the least mixing with the ICM, enabling the formation of long, coherent tails. The inviscid model has the shortest tail due to vigorous mixing, and the models with anisotropic viscosity are in between. The model with suppressed anisotropic viscosity due to plasma instabilities exhibits enhanced turbulence in the galactic tail and a concurrent limitation in viscous heating compared to the model neglecting plasma instabilities. These findings 
highlight the significant impact of ICM plasma physics on the processes of ram pressure stripping of galaxies.
\end{abstract}




\section{Introduction}

Ram-pressure stripping (RPS) happens when galaxies merge with galaxy clusters and interact with the intracluster medium (ICM). During this process, the infalling galaxies are subjected to a headwind exerting ram pressure on the interstellar medium (ISM), resulting in the stripped morphology of their tails (see \citealt{2022A&ARv..30....3B}, and references therein).

Observationally, many RPS galaxies exhibit disturbed tail structures. This morphology primarily arises from Kelvin-Helmholtz (KH) instability \citep{2014ApJ...784...75R}. The importance of viscosity can be estimated by the Reynolds number ($Re$), which is defined as the ratio of inertial forces to viscous forces. Because KH instability can be suppressed by the magnetic field and viscosity of the ICM, studying the morphology of RPS tails could provide constraints on the ICM magnetic field and viscosity. For instance, \citet{2015ApJ...806..104R} compared their simulation results with the observational data of Virgo galaxy M89 and found that the Reynolds number of the ICM should be around $Re \approx 50$, which corresponds to a $\sim 10\%$ suppression compared to the full isotropic Spitzer viscosity. Additionally, \cite{2024arXiv241107034I} analyzed the velocity structure function of the H$\alpha$ line in the tails of GASP jellyfish galaxies and found that the effective viscosity of the ICM is much lower than predicted by Coulomb collisions, measuring only $0.3\%-0.5\%$ of the expected Spitzer value. 

Previous simulations of the RPS process predominantly focused on investigating the composition \citep[e.g.,][]{2024arXiv240500768S} and stripping timescales \citep[e.g.,][]{2024arXiv240400129Z, 2024arXiv240402035G} of the RPS galactic tail, as well as examining the star formation induced by the RPS phenomenon \citep[e.g.][]{2023MNRAS.525.3551G, 2024arXiv240405676L}. Furthermore, investigations into the impact of ICM viscosity on RPS galaxies have thus far been confined to isotropic viscosity \citep{citation-key}. However, it is noteworthy that ICM viscosity is inherently anisotropic. Since the ICM is a weakly collisional, magnetized plasma, the ions can only gyrate around magnetic fields and propagate along their direction. Due to the low collisionality of the ICM plasma, pressure anisotropies develop according to the conservation of adiabatic invariants of particles. 
This pressure anisotropy acts like viscosity in the ICM, and its magnitude is determined by the balance between collisional relaxation and adiabatic production \citep{1965RvPP....1..205B, 2005ApJ...629..139S}. However, we found that the magnitude cannot exceed or go below certain values (see the Methods section for details); otherwise, plasma instabilities would be driven to reduce the pressure anisotropy \citep{2014PhRvL.112t5003K}, thereby limiting the magnitude of viscosity along the magnetic field lines.

As discussed previously, the simulations presented in \cite{citation-key} are confined to inviscid and isotropic viscosity scenarios, leading to the conclusion that ICM viscosity behaves akin to an inviscid fluid in the NGC4522 system. However, the authors suggested the presence of additional underlying processes that may be significant on a microscopic scale. The plasma instabilities mentioned above could potentially be the source of the suppression of ICM viscosity. For example, \cite{Kingsland_2019} investigated the evolution of active galactic nucleus (AGN) bubbles using 3D Braginskii-magnetohydrodynamic (MHD) simulations. They found that, even in the presence of anisotropic viscosity, the bubbles are eventually disrupted by the suppression of viscosity by plasma instabilities, similar to the inviscid case.
The results of \cite{Kingsland_2019} emphasize the importance of plasma instabilities in the development of hydrodynamic instabilities in the ICM. Therefore, it is our aim to study whether plasma instabilities can significantly influence the viscosity in the ICM and hence the morphology of RPS tails. 

In this paper, we focus on simulating four different ICM viscosity models to see how they affect the morphological evolution of galactic tails. Our models incorporate an inviscid ICM, an ICM with unsuppressed isotropic viscosity, an ICM with unsuppressed anisotropic viscosity, or an ICM with anisotropic viscosity suppressed by plasma instabilities. We expect the galaxy with isotropic viscosity to exhibit the brightest and most coherent tail, indicating minimal mixing with the ICM due to the suppression of KH instabilities. Conversely, the inviscid case is anticipated to be the most turbulent, resulting in the faintest tail due to extensive mixing driven by KH instabilities. The degree of mixing in the other two anisotropic viscosity models is expected to fall between these extremes. Whether plasma instabilities could suppress viscosity to a level similar to that in the inviscid case is one of the foci of our study.

The structure of this letter is as follows. In section \ref{section 2}, we summarize the simulation setup and the treatment of viscosity. In section \ref{section 3}, we present the stripping process of the galaxy and compare the four models. Finally, we conclude our findings in section \ref{section 4}.

\begin{table}
        \centering
	\caption{Simulation of four different ICM viscosity models}
	\label{tab:1}
	\begin{tabular}{c|c|c}
		Run ID & Model for ICM viscosity\\
            \hline
		(N) & no viscosity\\
		\hline
		  (I) & unsuppressed isotropic viscosity\\
            \hline
            (A) & unsuppressed anisotropic viscosity\\
            \hline
            (S) & anisotropic viscosity suppressed by plasma instabilities\\
		\hline
	\end{tabular}
\end{table}

\section{Methods}
\label{section 2}

We perform 3D Braginskii-magnetohydrodynamic simulations of a RPS galaxy using the FLASH code \citep{2000ApJS..131..273F}. The simulation setup of the initial ICM is identical to that in \citet[][]{2014ApJ...795..148T,2022MNRAS.512.5927F}, while the dimensions of our box differ. The simulations are performed in a box of dimensions with (xmin, xmax, ymin, ymax, zmin, zmax) = (-97.2 kpc, 97.2 kpc, -97.2 kpc, 97.2 kpc, -32.4 kpc, 162.0 kpc).
We model a disk galaxy centered at the origin (0,0,0) with the rotational axis aligned along the $z$-axis. The galaxy consists of a gaseous disk, a hot halo, a stellar disk and bulge, and a dark matter halo, initially in hydrodynamic equilibrium in the $z$ direction. The rotational velocity of the gas disk is determined to ensure that the combination of the centrifugal force and the pressure gradient of the disk balance the inward gravitational force (see \citealt[][]{2009ApJ...694..789T,2010ApJ...709.1203T,2022MNRAS.512.5927F}). The magnetic field within the galaxy follows a toroidal configuration. The field strength is set to be weak at the galactic center, peaking a few kpc from the center, and is assumed to be zero outside the disk. The rest of the simulation domain is filled with the ICM. We inject a headwind to blow in from the $-z$ boundary toward the $+z$ direction. The ICM wind is set with particle density $3\times10^{-4}\ {\rm cm}^{-3}$ (assuming a mean molecular weight of $\mu = 0.6$), temperature $7\times10^7$ K, and maximum wind speed 1300 km s$^{-1}$. The ICM gas density and temperature adopted in our model are similar to the observed values in the Coma Cluster at a distance of approximately $0.6$ Mpc from the cluster center \citep[][]{2013ApJ...775....4S,2014ApJ...784...75R}. The direction of the magnetic field vector is [1,1,0] and is perpendicular to the wind direction with a magnitude of $2\sqrt{2}\,\mu$G. We include a passively evolving tracer fluid, $f_{\rm gal}$, which represents the mass fraction of galactic gas within a grid cell ($0 \leq f_{\rm gal} \leq 1$). At $t=0$, we set $f_{\rm gal}=1$ in the galaxy and $f_{\rm gal}=0$ elsewhere. In the analyses shown in section \ref{section 3}, we define the tail region based on the selection criterion $0.01\leq f_{\rm gal} \leq 0.999$. The upper limit effectively excludes contributions from the galaxy, while the lower limit demonstrates low sensitivity to variations in its value. This criterion ensures that the selected range effectively identifies the stripped tail. 

In our simulation, we neglect the influence of radiative cooling and focus solely on the impact of viscosity. The treatment of viscosity in our simulations follows the methodology of \cite{Kingsland_2019}. Our simulations solve the Braginskii-MHD equations \citep[e.g.,][]{2015ApJ...798...90Z}
\begin{equation}
    \frac{\partial\rho}{\partial t}+\nabla\cdot(\rho\boldsymbol{\mathrm{v}})=0
\end{equation}
\begin{equation}
    \frac{\partial(\rho\boldsymbol{\mathrm{v}})}{\partial t}+\nabla\cdot\left(\rho\boldsymbol{\mathrm{v}}\boldsymbol{\mathrm{v}}-\frac{\boldsymbol{B}\boldsymbol{B}}{4\pi}+p_{\rm tot}\boldsymbol{I}\right)=\rho\boldsymbol{g}-\nabla\cdot\Pi
\end{equation}
\begin{equation}
    \frac{\partial E}{\partial t}+\nabla\cdot\left(\boldsymbol{\mathrm{v}}(E+p_{\rm tot})-\frac{\boldsymbol{B(v\cdot B)}}{4\pi}\right)=\rho\boldsymbol{g\cdot \mathrm{v}}-\nabla\cdot(\Pi\cdot\boldsymbol{\mathrm{v}})
\end{equation}
\begin{equation}
    \frac{\partial\boldsymbol{B}}{\partial t}+\nabla\cdot(\boldsymbol{\mathrm{v}B-B\mathrm{v}})=0,
\end{equation}
where $p_{\rm tot} = p + B^2/8\pi$ is the total pressure and $p$ is the gas pressure. 
The isotropic viscosity tensor \citep{spitzer2006physics} is defined as 
\begin{equation}
    \Pi_{\rm iso}=-\mu\nabla\boldsymbol{\mathrm{v}}
\end{equation}
and the anisotropic viscosity tensor is \citep{1965RvPP....1..205B}
\begin{equation}
    \Pi_{\rm aniso}=-3\mu(\boldsymbol{bb}-\frac{1}{3}\boldsymbol{I})(\boldsymbol{bb}-\frac{1}{3}\boldsymbol{I}):\nabla\boldsymbol{\mathrm{v}},
    \label{eq:6}
\end{equation}
where $\mu=2.2\,\times\,10^{-15}\,T^{5/2}\,/\,$ln$\Lambda$ g cm$^{-1}$ s$^{-1}$ is the dynamic viscosity coefficient (ln$\Lambda = 30$), and $\boldsymbol{b}$ is the magnetic field unit vector. 

In all the simulations, we set an upper limit for the kinematic viscosity coefficient $\nu = \mu/\rho = 5\times10^{30}$ cm$^2$ s$^{-1}$ to avoid simulation timesteps becoming too small and to save computational costs. This choice does not influence the main results of our study, because the expected values of $\nu$ remain below $\sim 10^{31}$ cm$^2$ s$^{-1}$ in most regions of the simulations. The main exceptions are the potential influence on the pressure anisotropies and the tail-ICM interface in case (I), where $\nu$ can exceed $10^{32}$ cm$^2$ s$^{-1}$ due to the high temperature gas generated by viscous heating. Consequently, the effect of the isotropy viscosity in case (I) may be underestimated. 



In Braginskii-magnetohydrodynamic, the viscosity originates from pressure anisotropy. Since the ICM is a weakly collisional, magnetized plasma, the ions propagate along the direction of the magnetic field and develop pressure anisotropy in order to conserve the adiabatic invariants. This pressure anisotropy acts like viscosity in the ICM. The pressure anisotropy can be described by the following equation \citep{2005ApJ...629..139S}
\begin{equation}
    p_\perp-p_\parallel=0.96\frac{p_\mathrm{i}}{v_{\mathrm{ii}}}\frac{\mathrm{d}}{\mathrm{dt}}\rm{ln} \frac{\emph{B}^3}{\rho^2}=3\mu(\boldsymbol{bb}-\frac{1}{3}\boldsymbol{I}):\nabla\boldsymbol{\mathrm{v}},
    \label{eq:7}
\end{equation}
where $p_\perp$ and $p_\parallel$ are the perpendicular and parallel components of the thermal pressure, respectively, $p_{\rm i}=\rho v^2_{\rm th}$ is the ion thermal pressure and $v_{\rm ii}=4\pi n_{\rm i} e^4 \ln \Lambda m_{\rm i}^{-1/2} T^{-3/2}$ is the ion-ion collisional frequency. The pressure anisotropy must be restricted to lie within a specific range
\citep{2014PhRvL.112t5003K}
\begin{equation}
    -\frac{2}{\beta}\leq\frac{p_\perp-p_\parallel}{p}\leq\frac{1}{\beta},
\end{equation}
where $\beta\equiv p/p_{\rm B}$ is the plasma beta (i.e., the ratio of the thermal pressure to the magnetic pressure). If pressure anisotropy violates this inequality, the plasma would become unstable to the firehose and mirror instabilities. At the saturated level, the resulting pressure anisotropies in the plasma would be pinned at the marginal stability thresholds. Since the effect of plasma instabilities has not been considered before in the context of RPS processes, we take it into account in the fourth model of our simulations. By combining Equations \ref{eq:6} and \ref{eq:7}, the anisotropic viscosity tensor can be expressed as
\begin{equation}
    \Pi_{\rm aniso}=-(\boldsymbol{bb}-\frac{1}{3}\boldsymbol{I}) p  \frac{p_\perp-p_\parallel}{p}.
\end{equation}
At the marginal stability bound,
\begin{equation}
    \Pi_{\rm aniso}=-(\boldsymbol{bb}-\frac{1}{3}\boldsymbol{I}) p
    \begin{cases}
    -\tfrac{2}{\beta}, & \text{if } \Delta p < -\tfrac{2}{\beta} \\
    \ \tfrac{1}{\beta}, & \text{if } \Delta p > \tfrac{1}{\beta}
    \end{cases}
\end{equation}

The four ICM viscosity models that we consider in our simulations are summarized in Table~\ref{tab:1}.

\section{Results and Discussion}
\label{section 3}
\subsection{Properties of the RPS tails}

\begin{figure}
    \centering
    \includegraphics[width=\columnwidth]{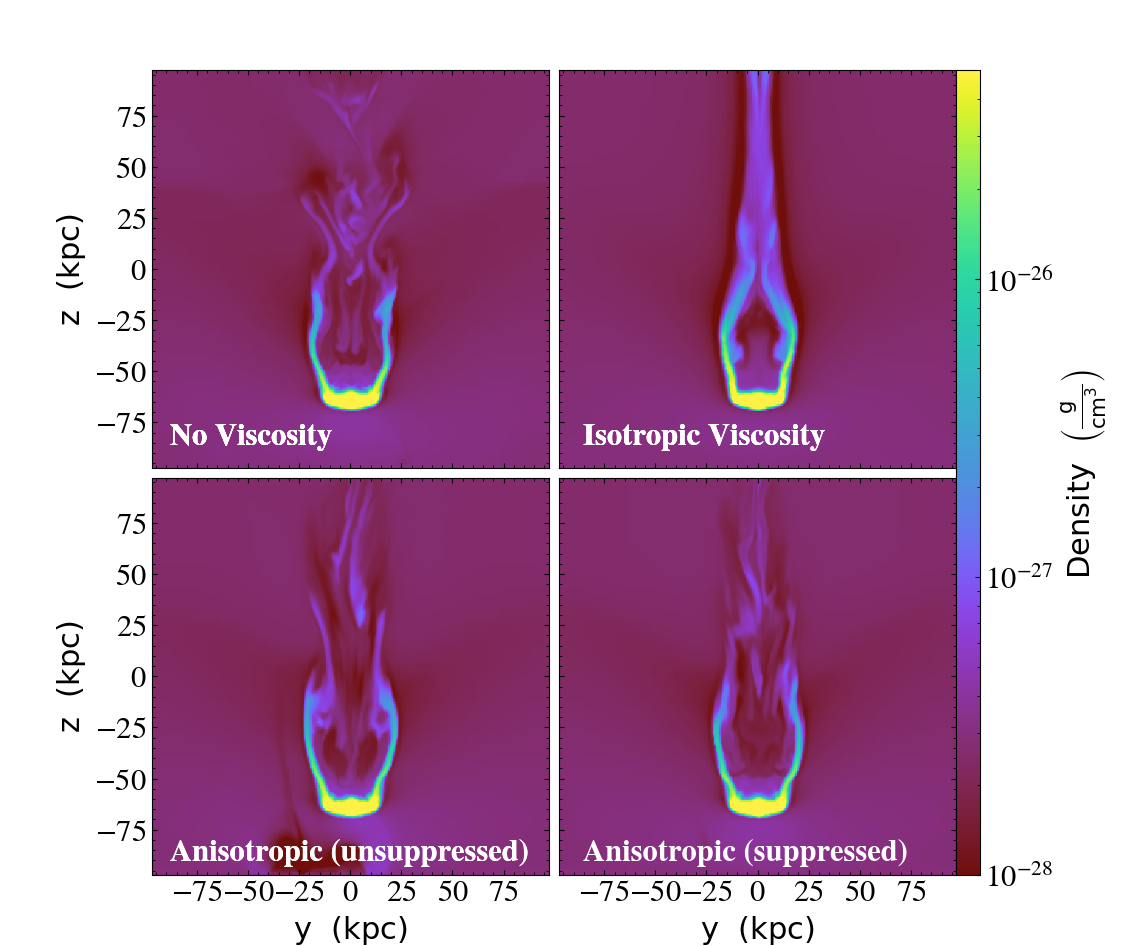}
    \caption{Density slice plots of four ICM viscosity models at $t=350$ Myr. The slice plot is cut through the plane $x=0$. The bifurcated tails resulting from the magnetic draping layer are clearly seen.}
    \label{fig:1}
\end{figure}

Our simulations model a disk galaxy encountering a headwind, resulting in RPS. 
Here we describe the overall evolution of the RPS process. Fig.~\ref{fig:1} presents density slice plots of the simulated galaxy along the $x=0$ plane, which lies between the directions parallel and perpendicular to the magnetic field, for the four viscosity models at $t=350$ Myr. The four models we consider (N), (I), (A), and (S) are defined in Table~\ref{tab:1}. The tails of the galaxy exhibit bifurcated appearance. This is attributed to the alignment of tail structures with the magnetic field directions. To further clarify this effect, we examined slices both along and perpendicular to the magnetic field direction (not shown here). Below $z=0$ kpc, the tails show little difference between the two views, indicating that the stripped galaxy exhibits a hollow-cylinder structure in this region. In contrast, above $z=0$ kpc, the tails bifurcate when viewed along the magnetic field but merge into a single tail when viewed perpendicular to it. This confirms previous findings that the magnetic field plays a key role in shaping the bifurcation structure of the galaxy \citep{2014ApJ...784...75R}. In addition, as found by \cite{2014ApJ...784...75R}, when the ICM wind interacts with a galaxy, the magnetic field would pile up on the side of the galaxy exposed to the incoming ICM wind and form a protective magnetic draping layer \citep[][]{2006MNRAS.373...73L,2008ApJ...677..993D}. 
The plasma beta in the draping layers decreases to approximately unity, indicating significant dynamical influence of the magnetic pressure. Due to magnetic draping, the magnetic field lines are wrapped around the galaxy. Magnetic pressure and magnetic tension in the draping layer then act to suppress hydrodynamic instabilities parallel to the field lines at the tail-ICM interface \citep[e.g.,][]{2018PhFl...30d4102L}, producing coherent double tails aligned with the ICM wind's magnetic field.

Initially, all four models display bifurcated tails shaped by the magnetic draping layer, with minimal differences among different physics cases. The stripped gas of the disk galaxy forms tails extended behind the galaxy. As time goes on, the tail becomes disturbed due to hydrodynamic instabilities at the tail-ICM interfaces (see Fig.~\ref{fig:1}). The perturbations subsequently grow and generate turbulence in the wake of the galaxy. 
The KH timescale is given by $t_{\mathrm{KH}} \sim 1.5\ \rm Myr\ (\frac{\mathrm{\lambda}}{3\ \rm kpc})(\frac{\Delta \mathrm{v}}{10^3 km\ s^{-1}})^{-1}\sqrt{\frac{0.1}{\eta}}$, where $\lambda$ is the perturbation wavelength, $\Delta \mathrm{v}$ is the velocity shear, and $\eta$ is the density contrast at the ICM-tail interface. The RT timescale is $t_{\mathrm{RT}} \sim 12.2\ \rm Myr\sqrt{\frac{1+\eta}{1-\eta}}\sqrt{\frac{10^{-8}}{g}}\sqrt{\frac{\mathrm{\lambda}}{3\ \rm kpc}}$, where $g$ is the gravitational acceleration in $cgs$ units. The perturbation wavelength $\lambda$ in our simulation is about 5 kpc. The density contrast $\eta$ at the ICM-tail interface and velocity shear $\Delta \mathrm{v}$ are determined from the density and z-velocity profiles cut along the y-axis at $x=0$ and $z=0$ with $\eta\sim1.09\times10^{-3}$ and $\Delta \mathrm{v}\sim 1.294\times10^3\ \rm km/s$. Based on the comparison between $t_{\mathrm{KH}}$ and simulation timescale ($t_{\mathrm{sim}}$), we find that $t_{\mathrm{KH}} \sim 18.5$ Myr is much shorter than $t_{\mathrm{sim}} \sim 450$ Myr, which indicates that KH instabilities are expected to develop within the simulation. Furthermore, we estimate $t_{\mathrm{RT}}$ to be approximately 59 Myr. Therefore, the order $t_{\mathrm{KH}} \ll t_{\mathrm{RT}} \ll t_{\mathrm{sim}}$ suggests that the hydrodynamic instabilities in our RPS simulation are predominantly governed by KH instabilities. %
The turbulence in the RPS tails acts to mix the dense, cold ISM with the diffuse, hot ICM wind. As the stripped gas mixes with the hot ICM, its properties, such as density and temperature, gradually change as a function of distance away from the galaxy. Over time, the mixing of the denser galactic gas with the less dense ICM reduces the overall density. Meanwhile, the temperature of the stripped tails can increase to around $10^7$ to $10^8$ K due to mixing with the hot ICM.

\begin{figure}
    \centering
    \includegraphics[width=\columnwidth]{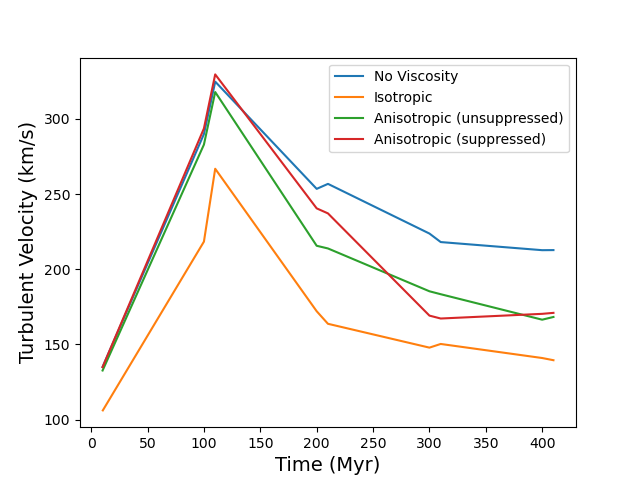}
    \caption{Evolution of the turbulent velocity of the four cases. Among the scenarios considered, the turbulent velocity is greatest for case (N), followed by case (S), (A), and lowest for case (I) for $t < 270$ Myr. This trend is consistent with the expectation that a higher level of viscosity would suppress KH instabilities and thus result in lower turbulent velocities.} 
    \label{fig:8}
\end{figure}

\begin{figure}
    \centering
    \includegraphics[width=\columnwidth]{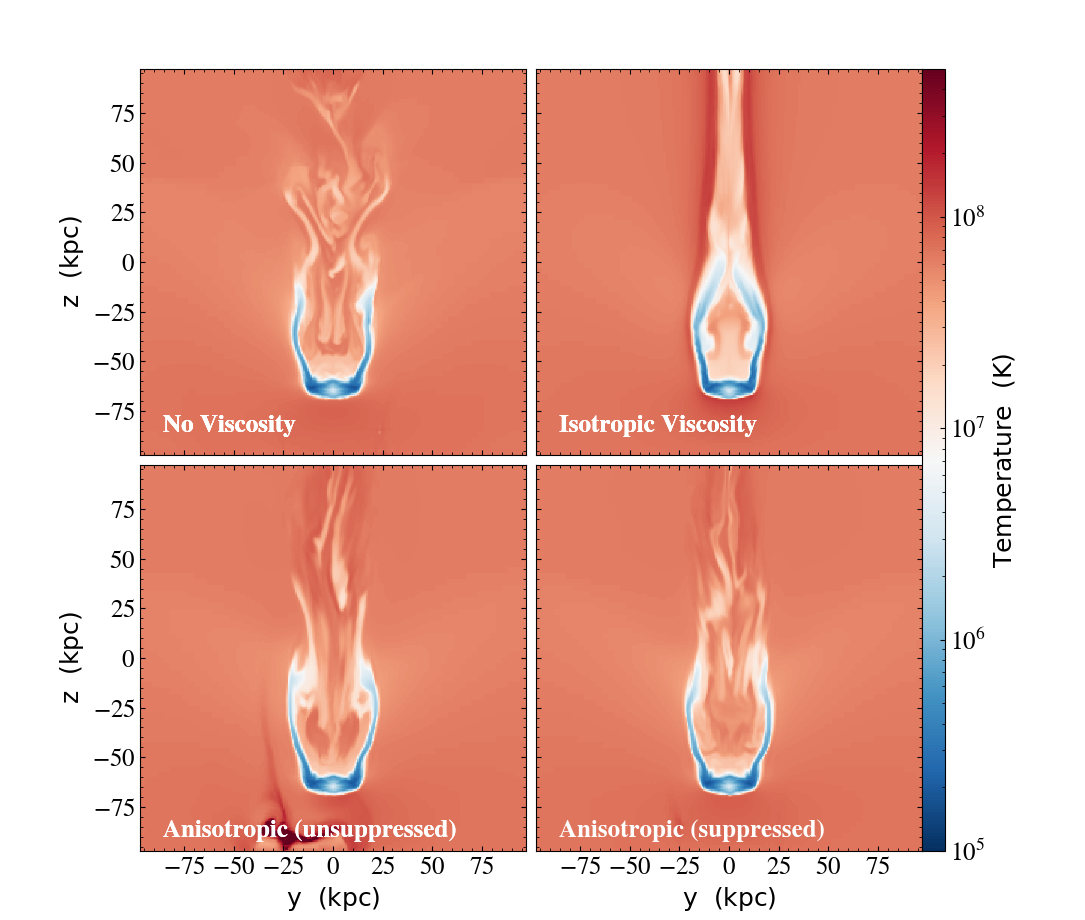}
    \caption{Slice plots of gas temperature of the four cases at $t = 350$ Myr. Case (I) exhibits the most pronounced viscous heating at the tail-ICM interface. Case (A) shows higher temperatures along its tails due to viscous heating compared to case (S).}
    \label{fig:4}
\end{figure}

\begin{figure}
    \centering
    \includegraphics[width=\columnwidth]{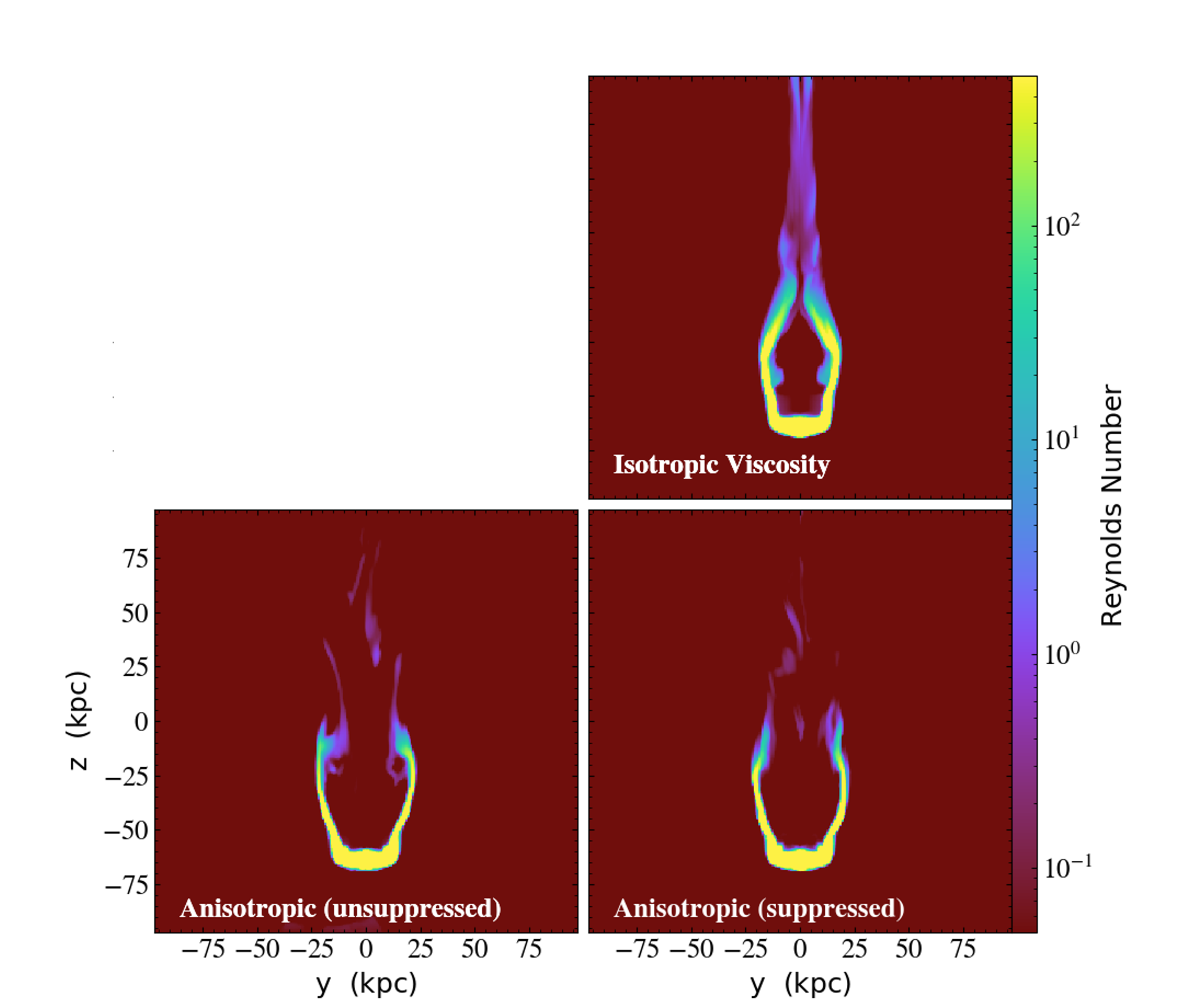}
    \caption{Slice plots of the Reynolds number ($Re$) for three cases at $t = 350$ Myr. Case (N) is omitted, as it has no viscosity and therefore no well-defined Reynolds number. For case (I) at about $z = -25$ kpc, $Re$ is the smallest among the four scenarios, indicating significant viscosity-driven suppression of KH instabilities. In contrast, the other three cases display insignificant differences in their $Re$ values, suggesting a limited impact of viscosity.}
    \label{fig:6}
\end{figure}

\subsection{Comparison of the four cases}
To investigate the effects of ICM viscosity on the morphology of RPS tails, we compare the differences among the four cases. Case (I) exhibits remarkably coherent tails in both slice plots of Fig.~\ref{fig:1}. This enhanced coherence is attributed to the effective suppression of KH instabilities by isotropic viscosity. In contrast, cases (N), (A), and (S) lack this suppression, leading to varying degrees of mixing with the ICM. While case (N) exhibits the shortest tail due to strong mixing, cases (A) and (S) show intermediate behavior. According to \cite{2021ApJ...911...68T}, the mixed fraction of the ICM increases as a function of height from the disk, further highlighting the significant mixing and the shortest tails observed in case (N). The level of mixing between the ISM and the ICM can be inferred from the level of turbulence in the RPS tails, which is shown in Fig.~\ref{fig:8}. Here we use $v_{\rm turb}=\sqrt{v_{\rm x}^2+v_{\rm y}^2}$ as a proxy for the turbulent velocity, as $v_{\rm z}$ is dominated by the bulk velocity of the stripped gas, and the x- and y-components of the bulk velocity are at the level of $\sim$20 km s$^{-1}$, which negligible compared to $v_{\rm turb}$. 
Among the scenarios considered, the turbulent velocity is greatest in case (N), followed by cases (S) and(A), and lowest for case (I) for $t < 270$ Myr with turbulent velocities in cases (N) and (S) being very similar before $t = 150$ Myr.
Higher level of turbulence in case (S) compared to (A) suggest that KH instability is not effectively suppressed when plasma instabilities reduce the level of viscosity.
For $t > 270$ Myr, the exact values are more sensitive to the chosen range of $f_{\rm gal}$. During this period, the separation between the level of turbulence between cases (N) and (I) becomes even more pronounced, whereas the differences between cases (A) and (S) diminish.

Further insights into the differences among the four cases can be obtained from the gas temperature slice plots. Fig.~\ref{fig:4} presents the gas temperature distributions at $t=350$ Myr. Notably, the galaxy tail in case (I) is the least turbulent among the four cases, consistent with the result in Fig.~\ref{fig:8}. Case (I) also exhibits the highest temperature at the interface between the tails and the ICM, indicative of significant viscous heating. Cases (A) and (S) are characterized by lower temperature, with case (N) showing the strongest extent of mixing and the lowest temperature. Additionally, case (S) exhibits lower temperature in its tails than case (A). This temperature reduction in case (S) is attributed to the suppression of anisotropic viscosity by plasma instabilities, which limits the extent of viscous heating.

To better understand the role of viscosity in shaping the morphology of the galaxy's tails, we analyzed the Reynolds number ($Re \equiv UL/\nu$). $Re$ is defined as the ratio of inertial forces to viscous forces, where $U\approx 4\times10^{7}$ cm s$^{-1}$ is the characteristic flow speed, $L=5$ kpc is the characteristic length scale, and $\nu$ is the kinematic viscosity. The slice plot of $Re$ is shown in Fig.~\ref{fig:6}. For case (I), $Re$ at the interface between the ICM wind and the galactic tail is the smallest ($Re\leq 1$) among the four scenarios, consistent with the strong viscous heating observed in Fig.~\ref{fig:4} and the significant impact of viscosity on suppressing KH instabilities. In contrast, the other two cases exhibit relatively minor differences and have larger $Re$ values, consistent with the limited impact of viscosity and the more turbulent tail morphology in these three cases.

\begin{figure}
    \centering
    \includegraphics[width=\columnwidth]{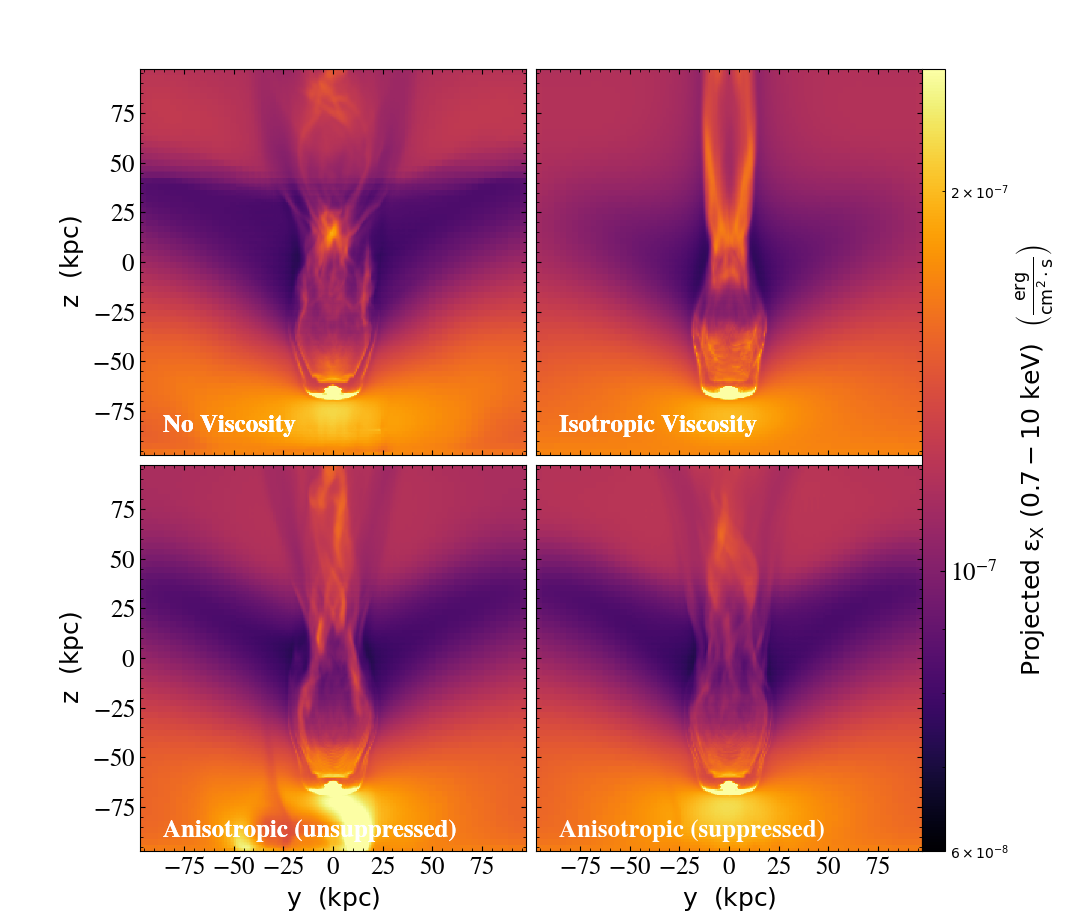}
    \caption{Simulated X-ray images in the 0.7-10 keV band at $t=350$ Myr for each model. The upper left, upper right, bottom left, and bottom right panels show the (N), (I), (A), and (S) cases, respectively. Among these, case (I) exhibits the highest brightness, followed by case (S), with cases (A) and (N) appearing comparatively fainter.}
    \label{fig:2}
\end{figure}

Finally, to further investigate the extent of mixing in each scenario, we present simulated X-ray images obtained by integrating the X-ray emissivity along the $x$-axis at $t = 350$ Myr (Fig.~\ref{fig:2}). 
Consistent with our expectations regarding the order of mixing among the four cases, case (I) exhibits the highest X-ray brightness due to the higher density of the tails. Although the differences among the remaining three cases are less pronounced in Fig.~\ref{fig:2}, case (A) shows slightly brighter X-ray emission than case (S) and case (N) in the tails, especially at the tail-ICM interface between $z = -30$ and $z = 0$ kpc. 
For $z > 0$ kpc, the tails in case (A) also appear brighter than those in case (S). These results support the interpretation that plasma instabilities effectively suppress viscosity, thus enhancing mixing and leading to fainter tails in case (S).


\subsection{Impact of the plasma instabilities}
The analysis in previous sections shows that the degree of mixing is highest in case (N), then (S), then (A), and lowest in (I), indicating the influence of the viscosity models.
In particular, the difference between the two anisotropic viscosity cases (A) and (S) is attributed to the influence of plasma instabilities, which we will explore in detail in this section.

\begin{figure}
    \centering
    \includegraphics[width=\columnwidth]{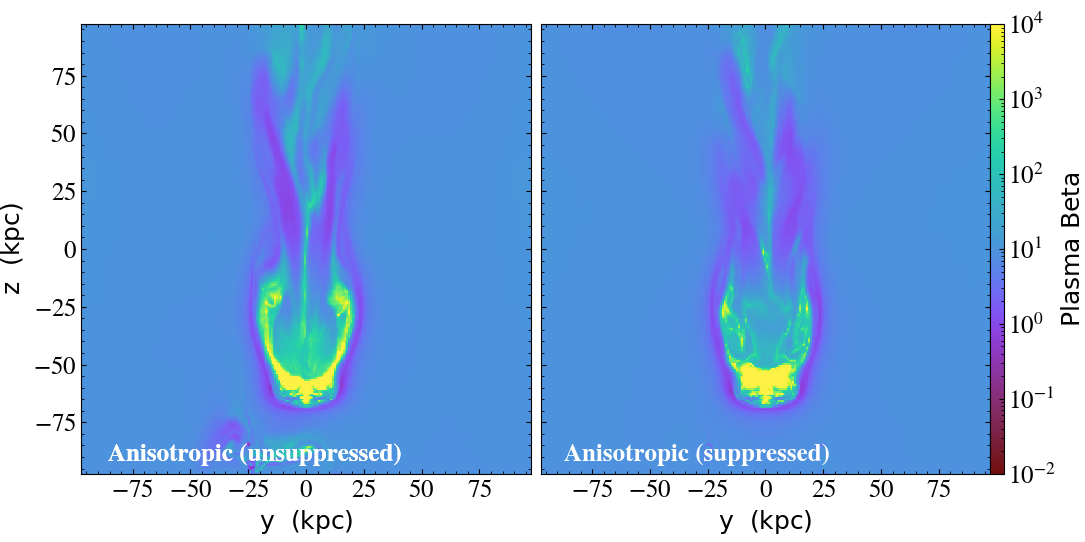}
    \caption{Slice plots of plasma beta ($\beta$) for the cases (A) and (S) at $t = 350$ Myr. The results indicate that between $z = -60$ and $z = -20$ kpc in cases (A) and (S), $\beta$ in the bifurcated tails of case (A) is a few times higher than in case (S).}
    \label{fig:5}
\end{figure}

The degree to which plasma instabilities affect the system is determined by the magnitude of plasma $\beta$ since the bound for pressure anisotropy is inversely proportional to $\beta$. Therefore, we examine plasma beta $\beta$ slice plots in Fig.~\ref{fig:5}. 
Between $z = -60$ and $z = -20$ kpc in cases (A) and (S), $\beta$ in the bifurcated tails of case (A) is a few times higher than in case (S). This can be attributed to stronger mixing between the RPS tails and the ICM, where turbulence amplifies the magnetic field and lowers $\beta$. The lower $\beta$ in case (S) therefore indicates enhanced mixing due to viscosity suppression by plasma instabilities. Since $\beta$ within the galaxies of both cases is large, magnetic pressure is dynamically unimportant, and the difference in $\beta$ between the anisotropic cases arises primarily from the mixing described above.


To further quantify the suppression of plasma instabilities, we also analyze a parameter that quantifies the departure from the marginal stability threshold, which is defined as $f_{\rm p} = \beta(\frac{p_\perp-p_\parallel}{p}) = \frac{\Delta_{\rm p}}{1/\beta}$. 
We find that, in case (A), the pressure anisotropy $\Delta_{\rm p}$ in the ambient ICM could reach values around $20-30$. 
This pressure anisotropy exceeds the marginal-stability threshold, which indicates that the system should be unstable to the plasma instabilities. When the bounds for pressure anisotropies are applied, the permitted range of pressure anisotropies is significantly constrained. As a result, $f_{\rm p}$ approaches 1 in ambient ICM in case (S). This indicates that plasma instabilities effectively suppress the viscosity by a factor of approximately $20-30$ compared to case (A). This could explain why the properties of RPS tails in case (S) are similar to those in the inviscid case (N). 

\section{Conclusions}
\label{section 4}
We conducted 3D Braginskii-MHD simulations to investigate the effects of viscosity on the morphological evolution of galactic tails, employing four distinct viscosity scenarios: inviscid (N), unsuppressed isotropic viscosity (I), unsuppressed anisotropic viscosity (A), and anisotropic viscosity bounded by plasma instabilities (S). Our findings yield the following conclusions:
\begin{enumerate}[(1)]
    \item Model (I) demonstrates the most coherent tails, suggesting the least mixing with the ICM, whereas model (N) shows the faintest and most turbulent tail. Models (A) and (S) fall in between these extremes. Based on the turbulent velocity evolution, model (S) exhibits greater turbulence compared to model (A), owing to the suppression of viscosity by plasma instabilities.\\
    \item Simulated X-ray images analysis reveals that case (I) has the highest brightness, followed by case (S), with cases (A) and (N) appearing comparatively fainter.\\
    \item Model (S) exhibits characteristics closer to model (N).
    This indicates the effective suppression of viscosity by plasma instabilities in case (S). Consequently, the viscosity in case (S) is diminished, leading to behavior resembling case (N).\\ 
    \item The analysis of $f_p$ (degree of departure from marginal stability threshold for plasma instabilities) indicates that the plasma instabilities effectively suppress viscosity in case (S) by a factor of $\sim 20-30$ compared to case (A). This is consistent with the trends of tail morphology found in our simulations -- with plasma instabilities, the tails are as turbulent as in the inviscid case (N).\\ 
\end{enumerate}

Plasma instabilities have recently become a vibrant and evolving area of research, with many questions still to be explored. Our study suggests that plasma instabilities could limit the pressure anisotropies in the ICM and thus provide additional suppression of the ICM viscosity around RPS tails. This offers a foundational step in understanding the role of plasma instabilities on the galactic RPS tails, shedding light on the intricate interplay between these instabilities and the galactic environment. While our initial model has provided valuable insights, future studies incorporating additional factors, such as varying inclination angles of the galaxy, radiative cooling, cosmic rays, and AGN feedback will be crucial for a more comprehensive understanding of the RPS process. 

\section*{Acknowledgements}
The authors thank Tsung-Chi Chen for his contribution for the initial setup of the simulation. YHT and HYKY acknowledge the support from the National Science and Technology Council (NSTC) of Taiwan (MOST 109-2112-M-007-037-MY3; NSTC 112-2628-M-007-003-MY3). HYKY acknowledges the support of the Yushan Scholar Program of the Ministry of Education (MoE) of Taiwan. This work used high-performance computing facilities operated by Center for Informatics and Computational in Astronomy (CICA) at NTHU. FLASH was developed in part by the DOE NASA- and DOE office of Science-supported Flash Center for Computational Science at the University of Chicago and the University of Rochester. MR acknowledges support from the National Science Foundation Collaborative Research Grant NSF AST-2009227.


\bibliographystyle{mnras}
\bibliography{example}


\bsp	
\label{lastpage}
\end{document}